\documentclass[aps,prx,twocolumn,twoside,floatfix,a4,showpacs,superscriptaddress]{revtex4-1}

\usepackage{amsmath, amssymb, amsfonts}
\usepackage{graphics, graphicx}
\usepackage{graphicx, epstopdf, epsfig}
\usepackage{dsfont}
\usepackage{color}
\usepackage{soul}

\usepackage{setspace}

\begin{document}	

\title{Logical Error Rate Scaling of the Toric Code}
\author{Fern H.E. Watson}
\email{fern.watson10@imperial.ac.uk}
\affiliation{Department of Physics, Imperial College London, Prince Consort Road, London SW7 2AZ, UK.}
\author{Sean D. Barrett}
\thanks{Deceased 19 October 2012.}
\affiliation{Department of Physics, Imperial College London, Prince Consort Road, London SW7 2AZ, UK.}

\begin{abstract}
To date, a great deal of attention has focused on characterizing the performance of quantum error correcting codes via their thresholds, the maximum correctable physical error rate for a given noise model and decoding strategy. Practical quantum computers will necessarily operate below these thresholds meaning that other performance indicators become important. In this work we consider the scaling of the logical error rate of the toric code and demonstrate how, in turn, this may be used to calculate a key performance indicator. We use a perfect matching decoding algorithm to find the scaling of the logical error rate and find two distinct operating regimes. The first regime admits a universal scaling analysis due to a mapping to a statistical physics model. The second regime characterizes the behavior in the limit of small physical error rate and can be understood by counting the error configurations leading to the failure of the decoder. We present a conjecture for the ranges of validity of these two regimes and use them to quantify the overhead -- the total number of physical qubits required to perform error correction.
\end{abstract}


\maketitle

\section{Introduction} 

Quantum computers are sensitive to the effects of noise due to unwanted interactions with the environment. To overcome this, fault-tolerant protocols that utilize error correction codes have been developed. These schemes allow arbitrary quantum gates to be performed in spite of the noise that is ubiquitous in current models of quantum computing. 

Recent progress has been made towards experimental implementations of quantum error correcting codes using small numbers of qubits realized using photonic systems, trapped ions and NMR techniques \cite{Chen2007, Lu2009, Yao2012, Schindler2011, Zhang2011}. Superconducting qubits are another promising experimental technique for scalable fault-tolerant quantum computing \cite{Paik2011, Rigetti2012, Nigg2013}, including surface code architectures \cite{DiVincenzo2009}.

The surface code \cite{Kitaev1997, Bravyi1998} is one of a family of topological codes, and is the basis for an approach to fault-tolerant quantum computing for which high thresholds have been reported \cite{Raussendorf2007, Raussendorf2007a, Barrett2010, Fowler2012}. The toric code \cite{Dennis2002} is among the most extensively studied of this family of codes, revealing much insight into related topologically ordered systems. A great deal of work has concentrated on calculating thresholds for various error models \cite{Stace2009, Wang, Hutter2013}, and on the discovery and implementation of new classical decoding algorithms \cite{Harrington2004, Duclos-Cianci2010, Duclos-Cianci2010a, Bombin2012, Wootton2012, Hutter2013, Fowler2013}. The toric code performs well, with high thresholds for some commonly studied noise models. 

A high threshold is a very desirable property of an error correcting code since for all error rates below the threshold, increasing the number of physical qubits encoding the quantum information reduces the logical error rate. In a realistic setting the code must be operating at an error rate below the threshold. Other quantities then become important to characterize the performance of a quantum computer, for example the {\em code overhead}, the number of physical qubits comprising the code that are required to adequately protect the encoded quantum information. This is an important consideration for the practical implementation of fault-tolerant quantum computation and has recently begun to draw some attention \cite{Bravyi2013, Gottesman2013, Suchara2013}.

The logical failure rate of the error correction, denoted here as $P_{\mathrm{fail}}$, is a key metric of the performance of a code, since it describes the likelihood of failing to protect the encoded quantum information. In this work we seek the logical failure rate of the toric code for fixed code distance and physical error rate, $p$. The code distance is the minimum length of a string-like operator that has a non-trivial effect on the code space, and in the case of the toric code such operators have a length equal to the lattice size $L$.

The toric code is a simple model that is closely related to other, more physically realistic systems. We expect therefore that results for the logical error rate scaling of the toric code could be applied in a range of other physical systems -- most obviously the planar code (with open, rather than periodic, boundary conditions) and with noisy syndrome measurements. The techniques to determine the scaling of the logical error rate should be analogous although the numerics would be expected to differ from the toric code case \cite{Fowler2013a}. Furthermore, once the scaling has been determined it can be used to calculate the fault-tolerant overhead for the planar code using the methods presented in this paper.

Below the threshold, the logical failure rate of a topological code is expected to reduce exponentially as we increase the code distance \cite{Dennis2002}. Although the code performance improves rapidly with increasing $L$, in the lattice of the toric code the total number of physical qubits scales as $O(L^2)$. Manufacturing, storing, and manipulating resources with such a scaling is a non-trivial task with technology available at present. We should then ask not simply how large we can make the code, but how many physical qubits are required to achieve a desired error correction performance.

In order to answer this question, we examine the behavior of the toric code in the presence of uncorrelated bit-flip and phase-flip noise. We numerically simulate the error correction procedure and use this to find the failure rate as a function of the input parameters $L$ and $p$ and find two operating regimes. The first of these, which we will call the {\em universal scaling hypothesis}, extends ideas by Wang {\em et al.} \cite{Wang2003} and uses rescaling arguments based on a mapping to a well-studied model in statistical physics (the 2-dimensional random-bond Ising model, or RBIM). This approach provides a good estimate for $P_{\mathrm{fail}}$ when the error weight (the number of qubits an operator acts on non-trivially) is high and code distance is large.

Rescaling arguments apply in the thermodynamic limit, and close to criticality, where the correlation length of the RBIM also diverges and the appropriate length scale is the ratio of the lattice size to the correlation length, $L/\xi$. As $p$ decreases there is a point at which finite-size effects begin to dominate and we no longer expect the universal scaling hypothesis to apply. This limit corresponds to low physical error rates, as well as small lattices.

The second approach extends ideas by Raussendorf {\em et al.} \cite{Raussendorf2007} and Fowler {\em et al.} \cite{Fowler2012} to find an analytic expression for $P_{\mathrm{fail}}$ in the limit $p\rightarrow 0$. When the error weight is low and the code distance is small this expression gives a good estimate of the logical failure rate. We will refer to this as the {\em low} $p$ {\em expression}.

Although we know the limits in which each of these approaches is valid, we would like to make some quantitative statements about the range of parameters for which each is applicable. We shall present a heuristic argument for the range of  $L$ and $p$ for which each regime gives a good approximation to the numerical data. 

The structure of the paper is as follows. In Sec. \ref{sec:toric_code} we review the toric code and its properties. Readers familiar with this material may wish to skip to Sec. \ref{sec:US} which discusses the universal scaling regime, in which rescaling arguments are used to estimate the logical failure rate. Sec. \ref{sec:low_p} describes the regime in which finite-size effects dominate the logical failure rate and the failure rate is dominated by spanning errors. In Sec. \ref{sec:validity} we present our conjectures regarding the ranges of validity of each of the two regimes described. In Sec. \ref{sec:results} we use these results to demonstrate techniques to determine the overhead as a function of the single qubit error rate and the logical error rate. We conclude in Sec. \ref{sec:conclusion}. 

\section{The toric code}\label{sec:toric_code}

\subsection{Background}

\begin{figure}
\includegraphics[scale=0.32]{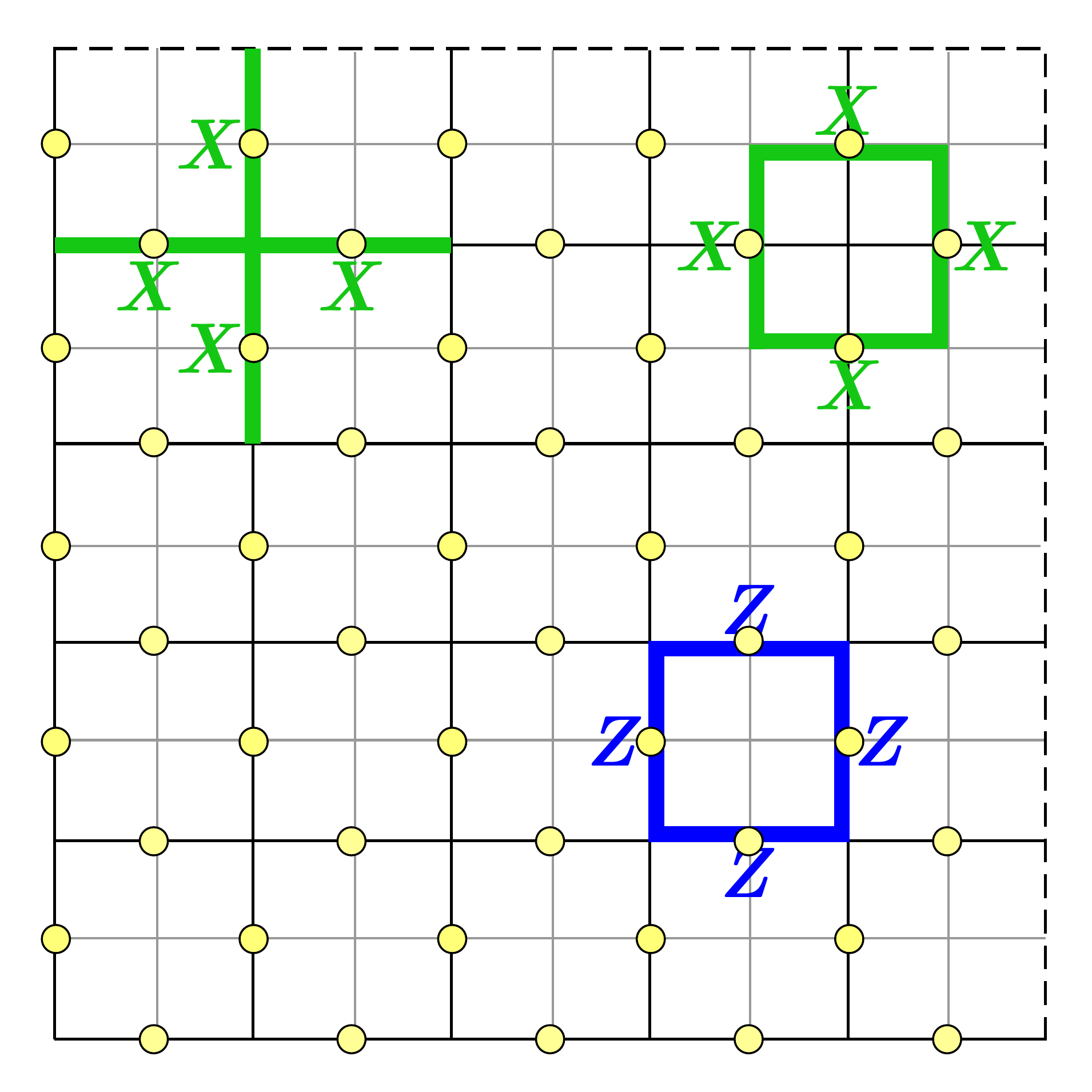} \centering
\caption{\label{fig:stabilizers}(color online). Representation of stabilizer generators on an $L = 5$ toric code lattice. Qubits, shown as yellow circles, are placed on the links of the lattice. Note that the periodic boundaries are indicated by the dashed lines. The dual lattice is shown using grey lines. {\em Top:} A vertex operator on the primal lattice (left) and the dual lattice (right). {\em Bottom:} A plaquette operator on the primal lattice.}
\end{figure}

In the toric code, physical qubits reside on the edges of an $L \times L$ square lattice, as shown in Fig. \ref{fig:stabilizers}. There are $n=2L^2$ physical qubits comprising the code. Periodic boundary conditions are imposed and the lattice can be imagined to be embedded on the surface of a torus. 

The toric code is described by a set of two types of commuting {\em stabilizer generators} --- the so-called {\em vertex}, $A_v$, and {\em plaquette}, $B_p$, operators, defined as 
\begin{eqnarray}
A_v = \displaystyle\otimes_{i \in v} X_i, \hspace{3mm} B_p = \displaystyle\otimes_{i \in p} Z_i,
\end{eqnarray}
where $X$ and $Z$ are the conventional single-qubit Pauli operators, $v$ indicates a vertex and $p$ a plaquette of the lattice. The $A_v$ operators therefore act on the four qubits surrounding a vertex of the lattice, and the $B_p$ operators act on the four qubits surrounding a plaquette, see Fig. \ref{fig:stabilizers}. These four-body measurements can be decomposed into four two-qubit CNOT gates with the addition of an ancilla \cite{Dennis2002}. 

We denote the logical encoded state of the toric code by $\left|\psi\right>_{\mathrm{toric}}$. In the absence of noise, measuring any element of $S = \left\{ A_v, B_p\right\}$ on this state will yield a $+1$ eigenvalue:
\begin{eqnarray}
S_i \left|\psi\right>_{\mathrm{toric}} = + \left|\psi\right>_{\mathrm{toric}},
\label{eqn:stabilized_state}
\end{eqnarray}
where $S_i \in S$. The stabilizer group is generated by $S$ with multiplication being the group action. All elements of the stabilizer group act trivially on the code space. The code-space of the toric code is four-dimensional and hence can encode two logical qubits. This is independent of $n$, hence the toric code protects a constant number of logical qubits regardless of its lattice size. 

The symmetry between the {\em primal lattice} and the {\em dual lattice} (constructed by replacing plaquettes of the primal lattice by vertices and vice versa) shown in Fig. \ref{fig:stabilizers}, reveals a useful symmetry in the stabilizers of the toric code. On the dual lattice the $A_v$ operators act on the qubits surrounding a plaquette, as shown in Fig. \ref{fig:stabilizers}. By considering both the primal and dual lattices we can view all stabilizers as closed loops, meaning that all plaquette-type operators on the primal lattice have an analogous vertex-type operator on the dual lattice. It follows that all results calculated for either bit-flip or phase-flip errors are interchangeable with results for the other type.

\begin{figure}
\includegraphics[scale=0.32]{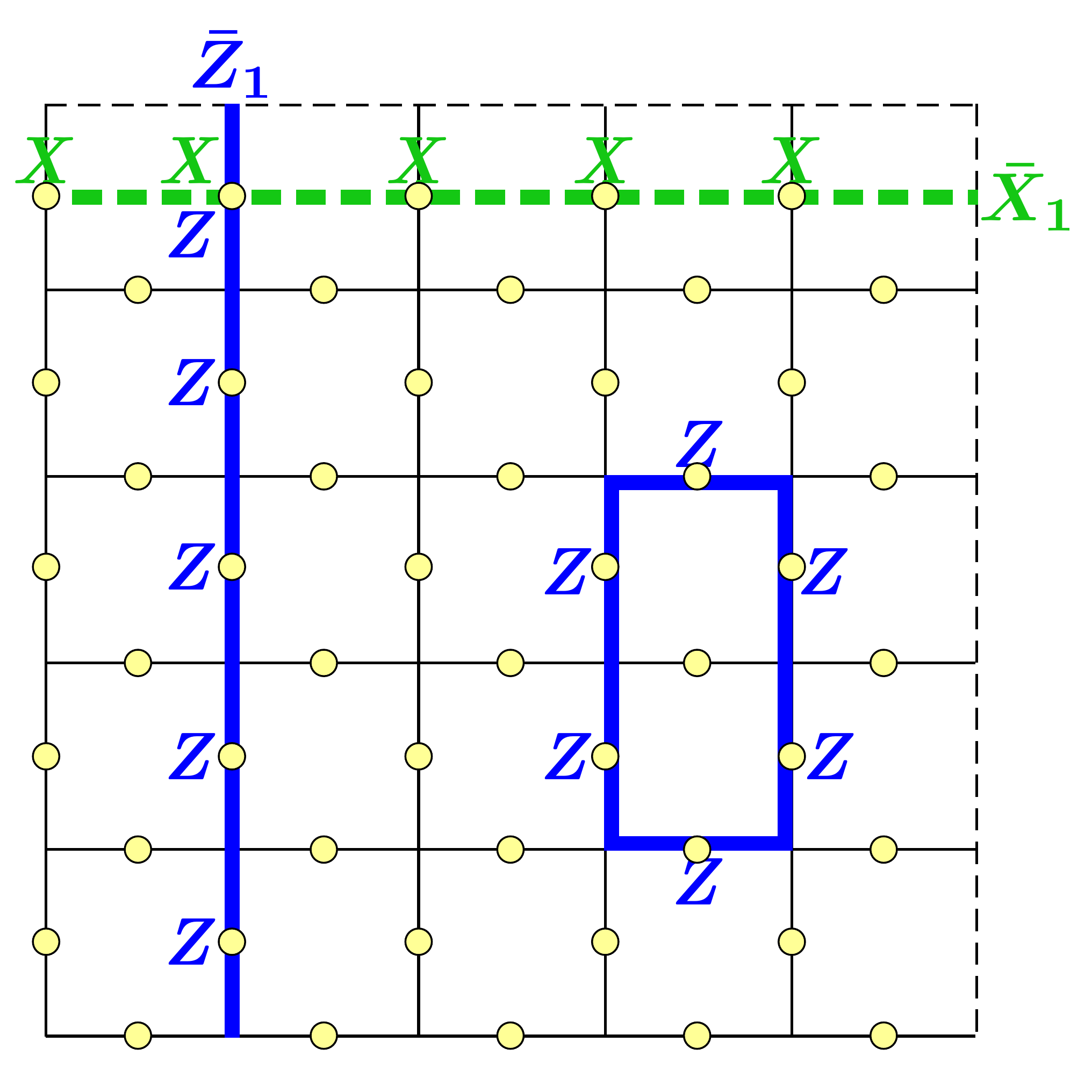} \centering
\caption{\label{fig:cycles}(color online). {\em Left:} $\bar{Z}_1$ is a minimum-weight homologically non-trivial cycle, equivalent to a logical operator acting on the encoded information. {\em Top:} The $\bar{X}_1$ operator, drawn as a cycle on the dual lattice (lattice not shown). The $\bar{X}_1$ logical operator shares a single physical qubit with $\bar{Z}_1$ and hence they anticommute. {\em Right:} An example of a homologically trivial cycle generated by multiplication of two adjacent plaquette operators.}
\end{figure}

In the language of algebraic topology, all of the stabilizers correspond to {\em homologically trivial cycles}. In Fig. \ref{fig:cycles} we show an example of a homologically trivial cycle that is generated by multiplying two adjacent stabilizer generators together. We see that all homologically trivial cycles act trivially on the code-space. 

The {\em logical operators} are also represented by cycles of Pauli operators. However, these cycles wrap around the torus and are not homologically equivalent to stabilizers. The logical operators correspond to {\em homologically non-trivial cycles} and have a non-trivial effect on the code-space. The minimum weight of a logical operator is $L$. 

There are two sets of $\bar{Z}$ and $\bar{X}$ logical operators addressing the two encoded qubits (overbar indicates a logical operation). One of these, labeled $\bar{Z}_1$, is shown in Fig. \ref{fig:cycles} spanning the lattice vertically. The corresponding $\bar{X}_1$ is also shown, and forms a closed horizontal loop on the dual lattice. By multiplying a logical operator by a subset of stabilizers we can continuously deform the minimum-weight cycle $\bar{Z}_1$ into any other operator spanning the lattice vertically. The set of operators that are equivalent up to stabilizer operations belong to the same {\em homology class} \cite{Hatcher2002}.

Errors are detectable if they anticommute with at least one element of the set of stabilizer generators $S$. In this work we assume that stabilizers are measured perfectly. It follows that if any non-trivial eigenvalues are observed, this indicates the presence of errors with certainty. The pattern of stabilizers that anticommute with a given error reveals some information about the location and most likely type of error, although it cannot uniquely identify the error. This ambiguity is due to the code degeneracy. 

\begin{figure}
\includegraphics[scale=0.32]{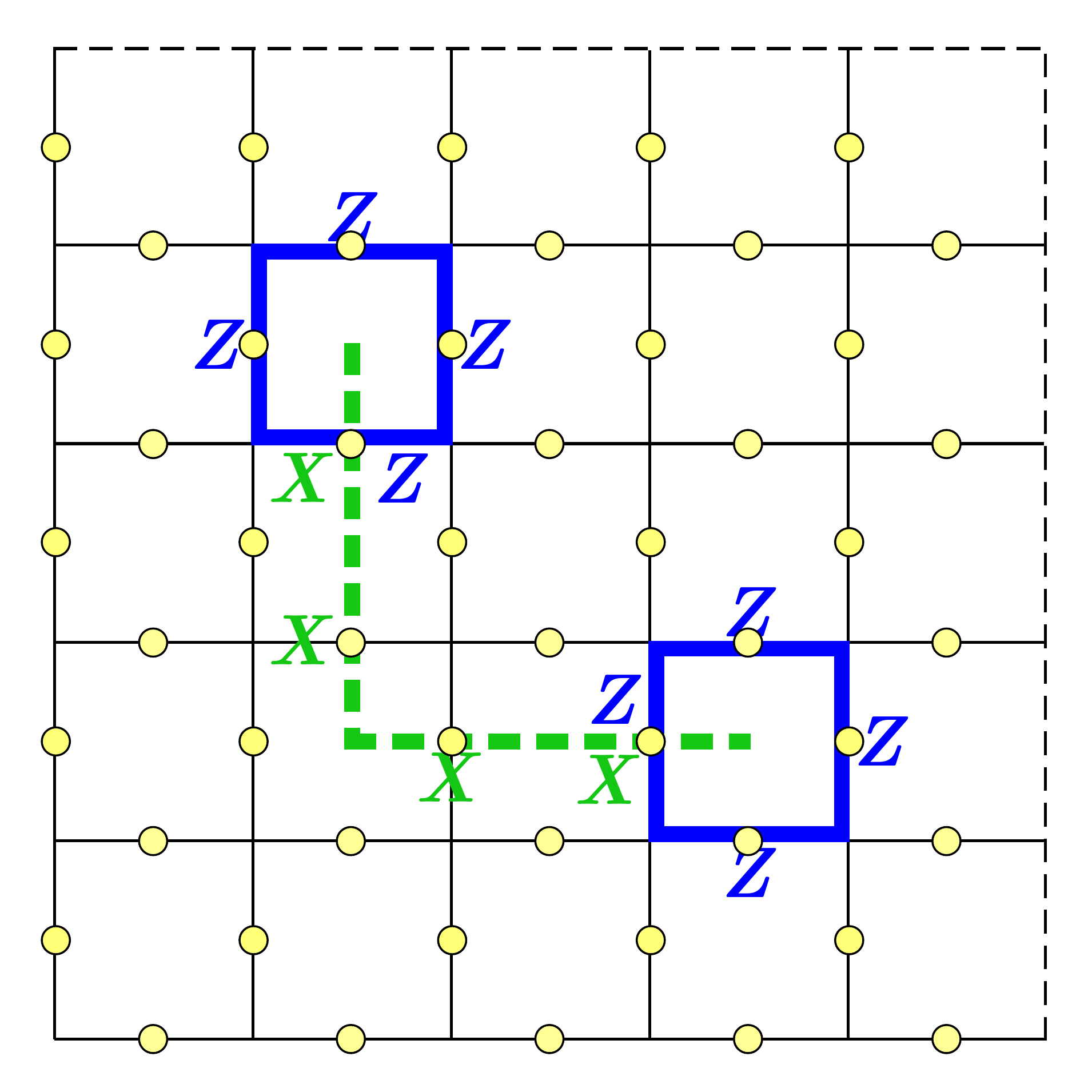} \centering
\caption{\label{fig:error_detection}(color online). A string of $X$ errors is shown as a dashed line on the dual lattice. Measuring the two $B_p$ generators indicated yields $-1$ eigenvalues because the stabilizer and error chain anticommute at these locations. Note that if the $X$ error chain forms a cycle then it will not be detectable.}
\end{figure}

The set of all errors on the lattice is called a {\em chain}, $E$. We use notation from algebraic topology to indicate the boundary of the chain of errors as $\partial E$. (A good introduction to algebraic topology can be found in many textbooks, for example see Ref. \cite{Henle1994}.) The errors commute with the stabilizers except at the boundary of the chain where the measured eigenvalues are non-trivial. The full set of stabilizer eigenvalues is called the {\em syndrome}. Fig. \ref{fig:error_detection} shows a string of $X$ errors and the two plaquette operators that anticommute with it. 

Once the syndrome has been established we employ a classical algorithm called a {\em decoder} to decide which correction chain, $E'$, to apply. The goal of the decoder is to pair the non-trivial syndromes such that the total operator $C=E+E^\prime$ has the highest probability of being a homologically trivial cycle and thus a member of the stabilizer group. Failure of the decoding algorithm corresponds to the creation of a homologically non-trivial cycle. The decoder used in this work, the minimum-weight perfect matching algorithm, is described in the next section.

\subsection{Error correction} \label{sec:error_correction}

The optimal threshold for the independent noise model that we consider here has been calculated using numerical techniques to be $p_{c} = 0.1093$ \cite{Honecker2001,Merz2002,Ohzeki2009,Queiroz2009}. However, there are no known {\em efficient} decoding algorithms that can obtain this threshold for the independent noise model on the toric code. 

Several classes of sub-optimal efficient decoding algorithm exist \cite{Duclos-Cianci2010, Bravyi2011, Wootton2012, Hutter2013}. The one used in this work is a version of Edmonds' minimum-weight perfect matching algorithm (MWPMA) \cite{Edmonds1965, Cook1999}. This algorithm pairs the non-trivial syndromes via a correction chain that has the least weight possible while satisfying the condition that its boundary matches the error chain boundary, i.e. $\partial E = \partial E^\prime$. This ensures that the total operator, $C=E+E'$, is a cycle. We denote the threshold for the MWPMA by $p_{c0}$. Numerical simulations suggest that $p_{c0} = 0.1031 \pm 0.0001$ \cite{Wang2003}.

Although this algorithm gives a high threshold \cite{Wang2003}, we shall consider a heuristic modification described in detail by Stace and Barrett \cite{Stace2010}, that includes the effects of the degeneracy of $E^\prime$ and can give thresholds up to $p_{c0} \approx 0.106$. Degeneracy counts the number of possible paths that the chain can take, given that its boundary and weight are fixed. Matchings with higher degeneracy have a higher probability of arising so they may be {\em a priori} more likely than some matchings with a lower weight. 

The degeneracy itself is simple to calculate for a given (minimum-weight) matching. For instance, for a path $m$ between two non-trivial syndromes, $a$ and $b$, the degeneracy of that path $D_m$ is given by the number of different combinations of the links in the matching. The product of all individual $D_m$ is the total degeneracy of the matching, $D_M$. 

To take degeneracy into account we compute the matching using the MWPMA, where the edge weights $d_{ab}$ are modified by the effect of the degeneracy of that path. Then the weight passed to the algorithm becomes $d_{ab} - \tau \ln D_m$. Here $\tau$ is a weighting that we assign to the degeneracy term. The degeneracy is added in such a way due to entropic considerations, see Ref. \cite{Stace2010} for details. The decoding algorithm minimizes this quantity globally and this has been shown to lead to an improved threshold \cite{Stace2010}. We refer to this enhanced version of the minimum-weight perfect matching simply as the PMA decoder. 

\subsection{Simulating noise and error correction}\label{sec:simulating_tc}

An important tool in this work is the numerical simulation of the detection and correction of errors on a toric code. Repeating random trials allows us to examine the failure probability of the code over a wide range of parameters. As stated earlier, we consider uncorrelated bit-flip and phase-flip errors arising at a rate $p$. It suffices to perform simulations for only one of these types of error since the results will be equivalent for the other.

The behavior of the toric code is simulated by placing an error with probability $p$ on each individual qubit of the toric code lattice of linear dimension $L$, giving rise to a (usually disjoint) error chain $E$. The syndromes are measured and the PMA decoder is used to determine the correction chain $E^\prime$. These correction chains are added, modulo 2, to $E$ and a parity check with each of the appropriate logical operators is used to determine the homology class of the total operator $C$. The result of this random sample indicates whether the error correction succeeds or fails.

The outcome of the Bernoulli trial (a single simulation of error correction) is assigned the value $n_{\mathrm{f}}=0$ if $C$ is in the trivial homology class and $n_{\mathrm{f}}=1$ if it is in any of the non-trivial homology classes. To gather statistics we repeat this procedure $N$ times for the same input parameters $(L,p)$. Of these $N$ trials, $N_{\mathrm{f}}=\sum_{i=1}^N n_{\mathrm{f},i}$ will have failed to perform error correction successfully. We therefore estimate the error correction failure probability as $P_{\mathrm{fail}}=N_{\mathrm{f}}/N$ and the variance of such a distribution is $\sigma^2 = P_{\mathrm{fail}}\left( 1-P_{\mathrm{fail}}\right)/N$. The resulting data $P_{\mathrm{fail}}(L,p)$ characterizes the toric code performance.

\section{The universal scaling hypothesis} \label{sec:US}

In Ref. \cite{Wang2003}, Wang {\em et al.} used ideas from the theory of critical phenomena in finite-sized systems to show that there is a critical point in the failure probability of the toric code. To do this, they used the 2-dimensional random-bond Ising model (RBIM) which is a model of ferromagnetism in which antiferromagnetic couplings arise at random. The probability distribution of antiferromagnetic couplings in this model matches the probability distribution of errors in the toric code, hence a mapping between the two models can be constructed \cite{Dennis2002, Wang2003, Harrington2004}. The RBIM has been extensively studied and it is known to undergo a phase transition from an ordered to a disordered phase as the concentration of antiferromagnetic bonds is increased. This implies a phase transition in the corresponding quantity of the toric code: its logical failure rate. 

Wang {\em et al.} demonstrated that for the regime where $L \gg \xi$,  where $\xi = (p-p_{c0})^{-\nu_{0}}$ is the RBIM {\em correlation length}, we expect scale-invariant behavior. This argument leads to the conjecture that in this regime the failure probability of the toric code is a function only of $L/\xi$ \cite{Wang2003}.

Below the threshold the failure rate is expected to depend exponentially on the system size \cite{Dennis2002, Raussendorf2007}, and also more generally in the fault-tolerant case \cite{Fowler2012a}.
\begin{eqnarray}
\ln P_{\mathrm{fail}} &\propto& -L.
\label{eqn:expL}
\end{eqnarray}
Numerical evidence for this will be provided later, in Fig. \ref{fig:exp_L}.

Together, the exponential dependence on $L$ and the scaling hypothesis fix the functional form of $P_{\mathrm{fail}}$.
\begin{eqnarray}
P_{\mathrm{fail}} &=& A e^{-a \left(L/\xi \right)} \label{eqn:US_hypoth}\\
&=& A e^{-a (p-p_{c0})^{\nu_{0}}L}.
\label{eqn:func_form}
\end{eqnarray}
In this expression $A$ and $a$ are constants that can be determined using numerical fitting techniques, see Sec. \ref{sec:testing_US_validity} and Appendix \ref{sec:threshold_calc}. 

In practice the toric code will be operating in the correctable ($p<p_{c0}$) regime so we use the rescaled variable $x=\left(L/\xi\right)^{1/\nu_0}$ (alternatively this may be written as $x=(p-p_{c0})L^{1/\nu_{0}}$) and we can rewrite the universal scaling hypothesis as
\begin{eqnarray}
P_{\mathrm{fail}} &=& A e^{-a x^{\nu_0}}. 
\end{eqnarray}

We determine the values of $A$, $p_{c0}$ and $\nu_0$ from a fit to data close to the threshold. In the remainder of this section we give evidence that the numerical data meets the two conditions required for the universal scaling hypothesis, namely an exponential decay of the failure rate as $L$ increases and scale invariance.

\subsection{Evidence for the universal scaling hypothesis} \label{sec:exp_L}

To observe the dependence of $P_{\mathrm{fail}}$ on $L$ and $p$ we have generated a set of Monte Carlo data for $0.01 \le p \le 0.08$ and odd lattice sizes in the range $5 \le L \le 23$. We use the simulation method outlined in Sec. \ref{sec:simulating_tc} with each simulation repeated $N=10^7$ times using Kolmogorov's Blossom V minimum-weight perfect matching algorithm implementation \cite{Kolmogorov2009}. We pass modified weights to the algorithm to account for degeneracy as described in Sec. \ref{sec:error_correction}. 

\begin{figure}
\includegraphics[scale=0.48]{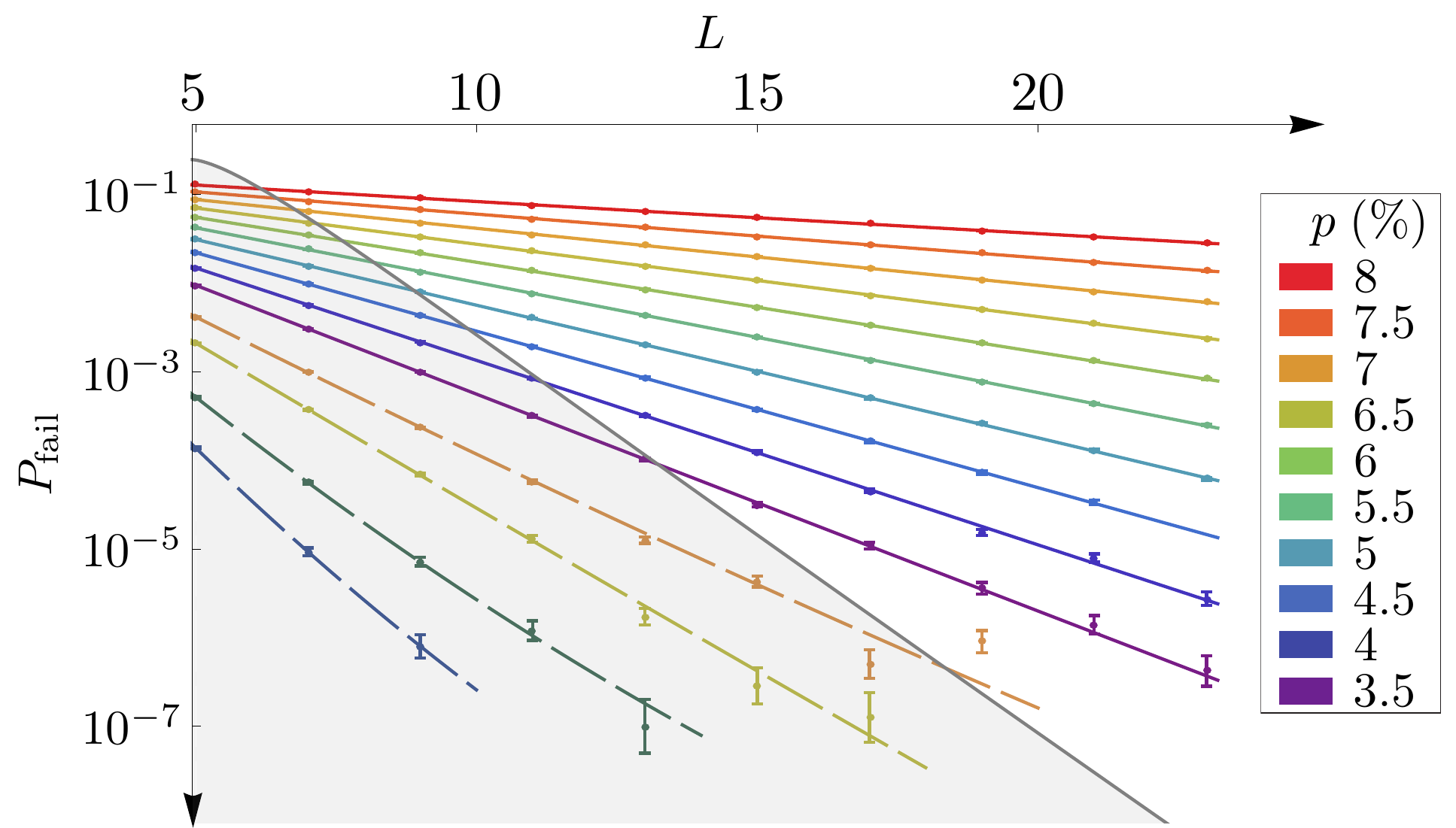} \centering
\caption{\label{fig:exp_L}(color online). Dependence of the logical failure rate $P_{\mathrm{fail}}$ on the size of the lattice. Each data point represents $N=10^7$ runs. The data is plotted on a logarithmic scale and linear fits to a selected set of the data between $p=3.5 \%$ and $p=8 \%$ are shown. The four data sets shown in in the lower part of the plot (dashed lines) are examples of data with $p<3.5 \%$ for which linear fits could not be identified. In the grey region the linear relationship is expected to break down according to our validity conjecture, see Sec. \ref{sec:validity}.}
\end{figure}

In Fig. \ref{fig:exp_L} we plot the logical failure rate on a logarithmic scale, as a function of the lattice size. The shaded portion of the figure indicates the region where this exponential relationship is not expected to hold according to a conjecture that will be explained in Sec. \ref{sec:validity}. 

Each set of data in Fig. \ref{fig:exp_L} is fitted using a quadratic ansatz in $L$:
\begin{eqnarray}
\ln P_{\mathrm{fail}} = \alpha + \beta L + \gamma L^2.
\label{eqn:quad_fit}
\end{eqnarray}
For data in the range $0.035 \le p \le 0.08$ and $5 \le L \le 23$ the quadratic coefficient $\gamma$ is typically 2--3 orders of magnitude smaller than the linear coefficient $\beta$. This is strong evidence for a linear fit to the (logarithmic) data, suggesting a fit of the form $P_{\mathrm{fail}} \propto e^{-L}$, matching equation (\ref{eqn:expL}). For data with values of $p<0.035$ the quadratic coefficient was comparable in magnitude to the linear coefficient. A selection of this data is also shown in Fig. \ref{fig:exp_L}, demonstrating that the behavior of the data for these values of physical error rate is ambiguous. Nevertheless, Fig. \ref{fig:exp_L} establishes an exponential dependence of the logical failure probability on $L$ for a wide range of the data. 

\begin{figure}
\includegraphics[scale=0.42]{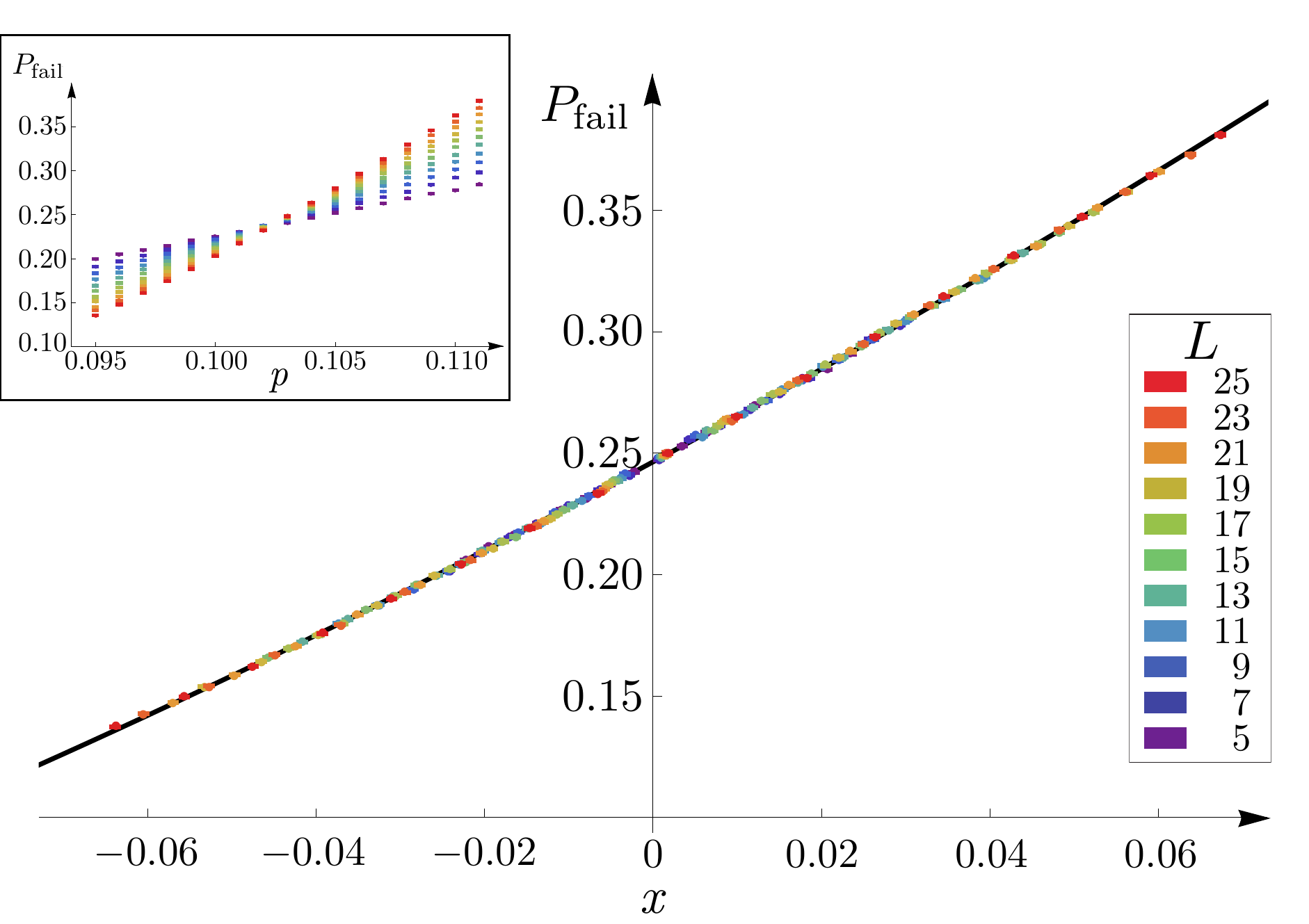} \centering
\caption{\label{fig:threshold}(color online). Data obtained from numerical simulations of the toric code failure rate close to threshold, rescaled using $x=(p-p_{c0})L^{1/\nu_0}$. Each data point represents $N=10^6$ runs. The finite-size correction $D L^{-1/\mu}$ is subtracted from $P_{\mathrm{fail}}$. All of the data collapses to a single curve and the threshold can be extracted as a fit parameter. {\em Inset:} The data prior to rescaling.}
\end{figure}

The universal scaling hypothesis in equation (\ref{eqn:US_hypoth}) also requires the system to be scale invariant which implies that the behaviour of $P_{\mathrm{fail}}$ should depend only on the length scale $L/\xi$. This is demonstrated in Fig. \ref{fig:threshold} which shows the results of numerical simulations of the toric code failure rate close to threshold. The plot will be explained in detail in Appendix \ref{sec:threshold_calc} but now we simply note that rescaling the numerical data using the variable $x = (L/\xi)^{1/\nu_0}$ leads to {\em data collapse}. This phenomenon describes the situation when data generated in different systems, in this case different lattice sizes, falls onto the same curve after an appropriate rescaling has been applied.

\section{The low single qubit error rate regime} \label{sec:low_p}

The universal scaling hypothesis is a good model for the logical failure rate when the lattice size is large and when there are sufficiently many errors. For a fixed lattice size, as $p$ is reduced the universal scaling behavior should not be expected to hold indefinitely. Indeed the numerical evidence suggests that when $p$ becomes sufficiently small the scaling hypothesis fails. In the $p\rightarrow 0$ limit the behavior is given by the low $p$ analytic approximation:
\begin{eqnarray}
P_{\mathrm{fail}} = \frac{2L \: L!}{\lceil L/2\rceil!\lfloor L/2\rfloor !} p^{\lceil L/2\rceil}.
\label{eqn:low_p_approx}
\end{eqnarray}

\begin{figure}
\includegraphics[scale=0.18]{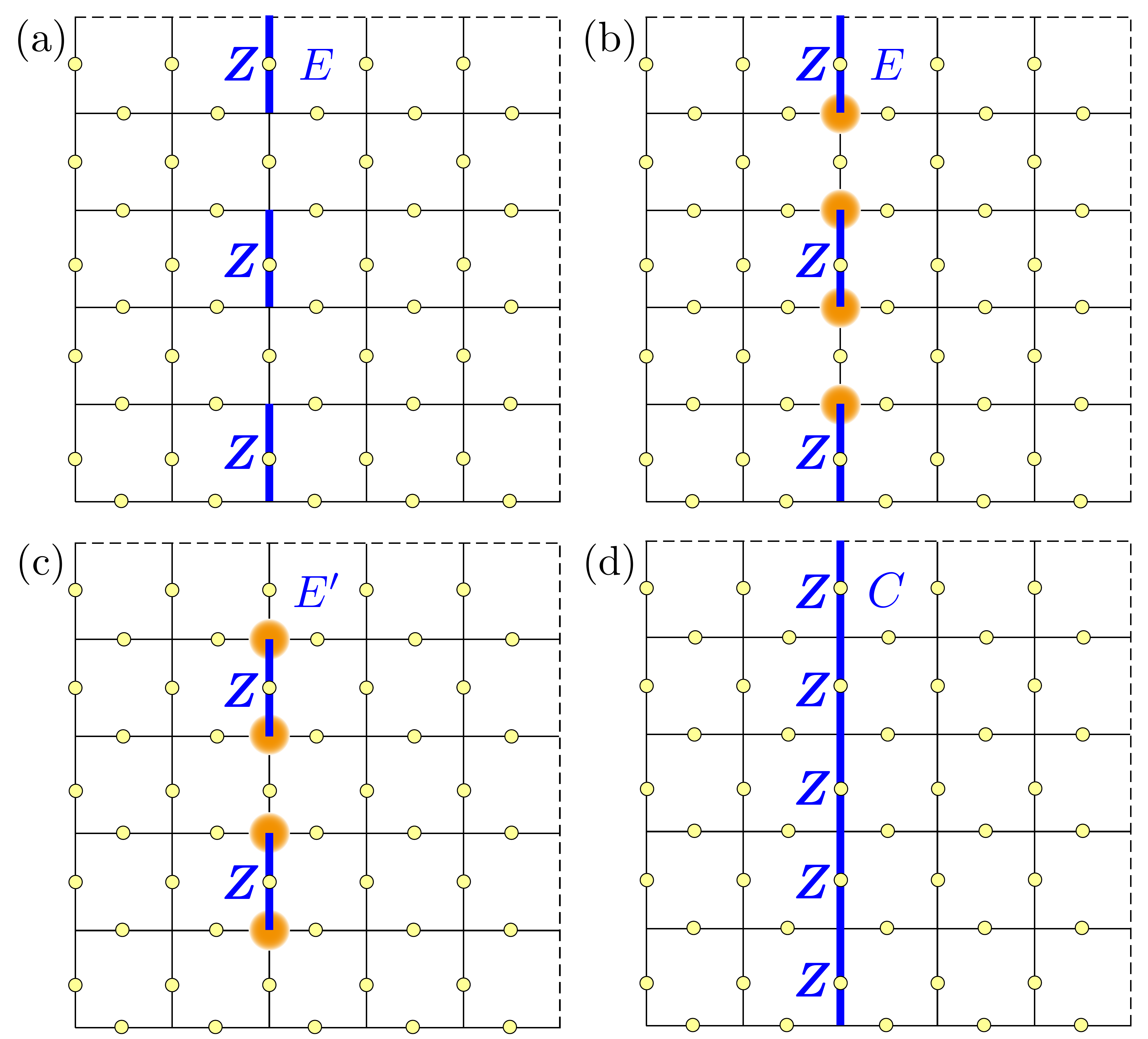} \centering
\caption{\label{fig:logical_fail}(color online). One way in which $\lceil L/2 \rceil$ errors lying along a minimum weight homologically non-trivial cycle will result in a logical error. The PMA decoder applies a correction chain that results in a non-trivial cycle, causing a logical failure. (a) The errors are distributed arbitrarily along one minimum-weight homologically non-trivial cycle of the lattice. (b) The syndromes that arise as a result of the error configuration are shown. (c) The minimum-weight perfect matching returns the correction chain $E'$ with certainty. (d) The resultant cycle $C=E+E'$ is homologically non-trivial, which means that the error correction has failed.}
\end{figure}

This is justified by considering the uncorrectable error configurations in the $p\rightarrow 0$ limit and calculating $P_{\mathrm{fail}}$ directly. Restricting ourselves to low single qubit error rates we consider the minimum number of errors that can cause the error correction to fail, $\lceil L/2\rceil$. To cause the error correction to fail these errors must lie along a single minimum-weight homologically non-trivial cycle of the toric code. If they fall in this way the PMA will certainly apply the remaining $\lfloor L/2\rfloor$ single qubit operators required to ensure $C=E+E'$ is a logical operator. Fig. \ref{fig:logical_fail} shows a sketch of how this happens.

Thus the expression in equation (\ref{eqn:low_p_approx}) for the failure rate is constructed via a counting argument. The first factor, $2L$, is the number of minimum-weight homologically non-trivial cycles of the code that exist. The second is the binomial coefficient which counts the possible combinations of $\lceil L/2\rceil$ errors along a cycle of weight $L$. Finally we include a factor that accounts for the likelihood of exactly $\lceil L/2\rceil$ errors occurring on a lattice constructed from $2L^2$ qubits, which is $p^{\lceil L/2\rceil} \left(1-p \right)^{2L^2 - \lceil L/2\rceil}$. The single qubit error rate is small so we can neglect the final factor of $\left(1-p \right)^{2L^2 - \lceil L/2\rceil}$ to obtain equation (\ref{eqn:low_p_approx}). In the low $p$ limit the $L$ dependence is $P_{\mathrm{fail}} \propto e^{-\lceil L/2\rceil}$ and we see that it is quantitatively different to the universal scaling regime, $P_{\mathrm{fail}} \propto e^{-L}$.

\section{The validity of the two regimes} \label{sec:validity}

The range of parameters we consider in our numerical simulations encompasses both the small $p$ limit and the universal scaling limit. For small single qubit error rates the weight of the errors is typically much smaller than the code distance and the low $p$ analytic expression is applicable. Conversely, for large $L$ the number of errors can be much larger than the code distance and we expect a universal scaling hypothesis to apply. These regimes are distinct, as we see from their differing dependence on the code distance. Each of the two regimes will provide a good approximation to the numerical data over some region of parameter space. We shall now make a heuristic argument to quantify those regions.

In order to make a conjecture about the validity of the regimes we consider the distribution of the number of errors that arise on a lattice of fixed size, at a known physical error rate. We will relate this distribution to $\lceil L/2 \rceil$, half the code distance. This number is significant to the PMA decoder because if the weight of the error chain, $|E|$, is less than this number then the error is certainly correctable. In the case when $|E| \ge \lceil L/2 \rceil$ a subset of the possible error configurations will lead to an incorrect pairing of syndromes, causing a logical failure. These are the spanning errors illustrated in Fig. \ref{fig:logical_fail}.

The typical weight of errors on the lattice can be shown to be $2L^2 p$. If $2L^2 p < \lceil L/2 \rceil$ then the expected number of errors is less than half the code distance and logical errors are dominated by spanning chains, see  Fig. \ref{fig:logical_fail}. For a fixed $p$, as $L$ increases this inequality is violated. When the number of errors is much greater than $L$ but they are typically correctable, this is the universal scaling limit.

Requiring $p\gg 1/L$ (up to a numerical factor) leads to a relationship between $L$ and $p$ that determines a minimum single qubit error rate for a given lattice size below which the universal scaling hypothesis breaks. We make the arbitrary but natural choice that the mean number of errors on the lattice must be two standard deviations above $\lceil L/2 \rceil$, leading to the expression
\begin{eqnarray}
p_{\textsc{ush}} \approx \frac{L^2+\sqrt{2 L^3}+2 L}{4 L^3}.
\label{eqn:p_min}
\end{eqnarray}
This expression, derived fully in Appendix \ref{sec:p_break}, determines whether the behavior can be considered to be within the universal scaling regime. 

We can find an equivalent expression for $p \ll 1/L$, when the single qubit error rate above which the low $p$ expression no longer provides a good approximation to the numerical data. This can be shown to be
\begin{eqnarray}
p_{\textsc{l}p} \approx \frac{L^2-\sqrt{2 L^3}+2 L}{4 L^3}.
\label{eqn:p_max}
\end{eqnarray}
When $p\sim 1/L$ there is a `crossover' region, in which the logical failure rate cannot be considered to be well approximated by either regime. 

\subsection{Testing the Range of Validity of the Universal Scaling Hypothesis}\label{sec:testing_US_validity}

\begin{figure}
\includegraphics[scale=0.43]{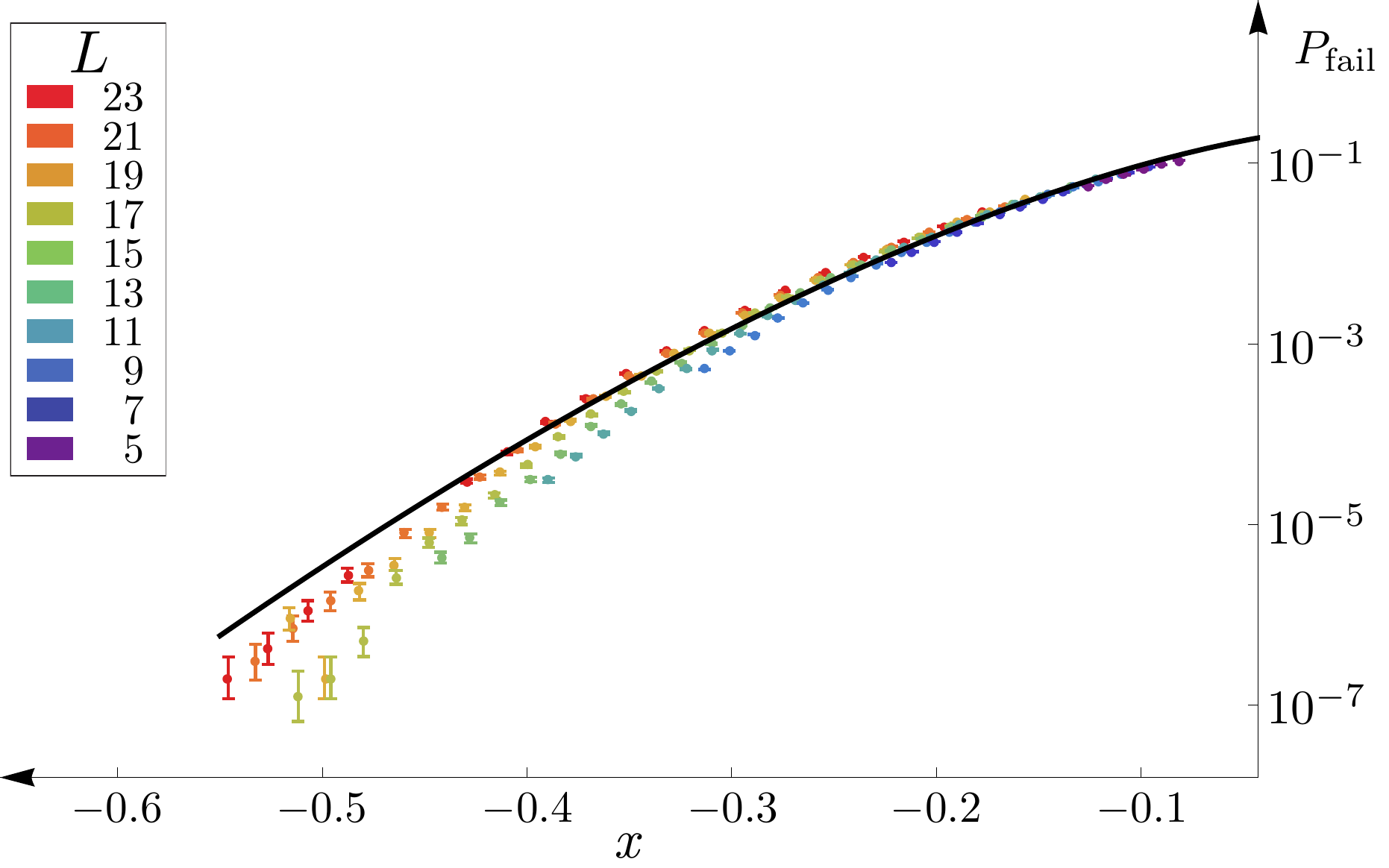} \centering
\caption{\label{fig:US_fit}(color online). Data satisfying the condition $p>p_{\textsc{ush}}$, plotted on a logarithmic scale and colored according to lattice size.  Each data point represents $N=10^7$ runs. Also shown in black is the fit of the ansatz, equation (\ref{eqn:US_hypoth}) with all values taken from the threshold fit (see Appendix \ref{sec:threshold_calc}) except for $a$ which was extracted using a fit to the data set shown.}
\end{figure}

Substituting $p_{\textsc{ush}}$ given by equation (\ref{eqn:p_min}) into the universal scaling hypothesis in equation (\ref{eqn:US_hypoth}) yields an expression for the minimum $P_{\mathrm{fail}}$, for a fixed $L$, that belongs to the universal scaling regime. This expression is plotted as a grey line in Fig. \ref{fig:exp_L} and hence the grey region indicates the region of parameter space where we do not expect the universal scaling hypothesis to hold. This supports the previous observation that most of the data we have obtained for $p<3.5 \%$ would lie outside the universal scaling region and therefore be poorly fit by equation (\ref{eqn:func_form}).

We have fitted the universal scaling ansatz, equation (\ref{eqn:US_hypoth}) to the data that falls outside this grey region. (The values of $A$, $p_{c0}$ and $\nu_0$ are all determined from the fit to the data around threshold.) From the fit to the data in the universal scaling regime we find $a = 32.31 \pm 0.13$. The data obeying the validity condition and the fit are shown in Fig. \ref{fig:US_fit}. 

\begin{figure}
\includegraphics[scale=0.46]{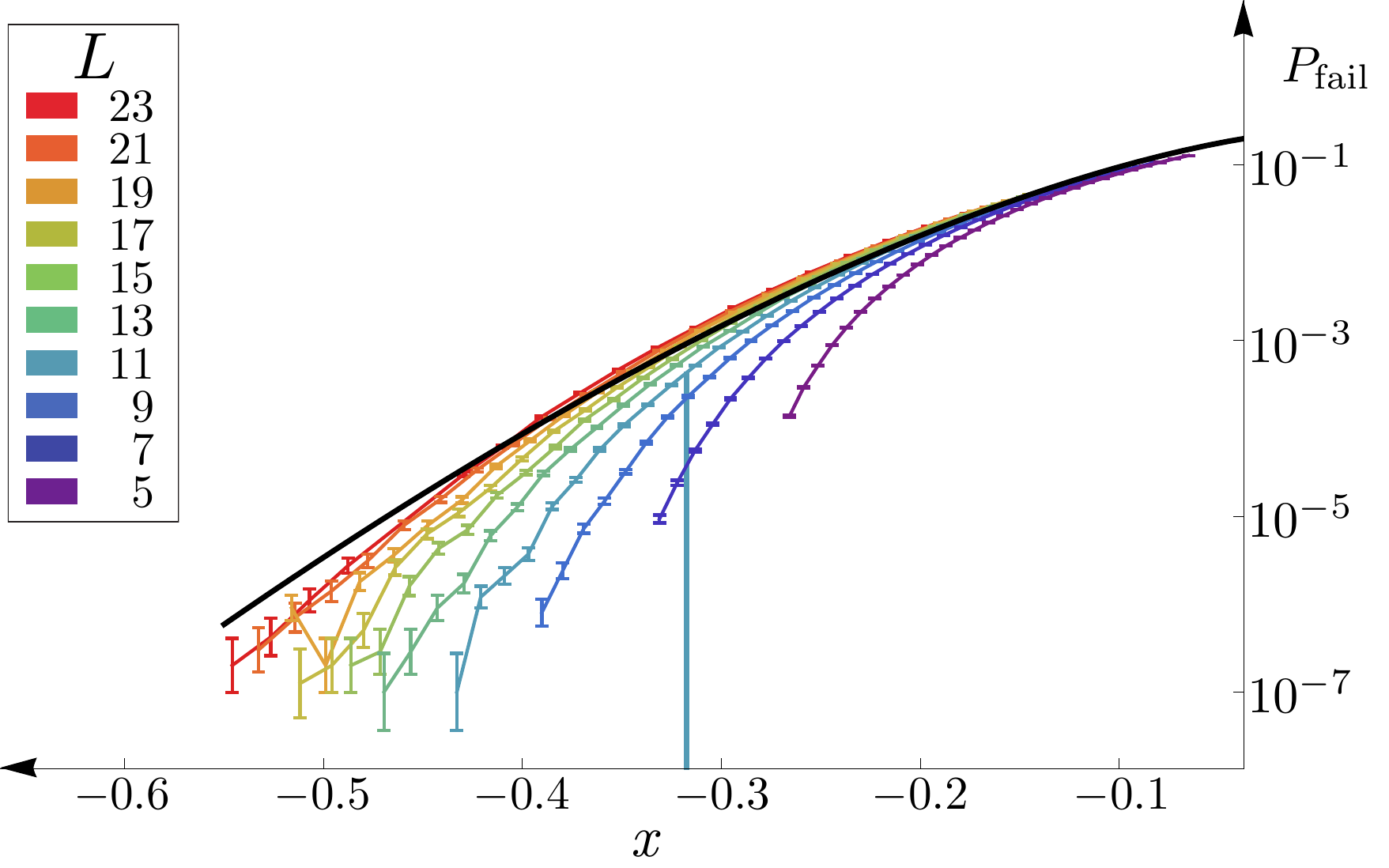} \centering
\caption{\label{fig:varyp}(color online). Logarithmic plot of all numerical data following the rescaling transformation $(L, p)\rightarrow x=(p-p_{c0})L^{1/\nu_0}$. The universal scaling fit is also shown in black. The data is plotted on a logarithmic scale and colored according to lattice size $L$. For fixed $L$, decreasing $x$ corresponds to reducing $p$. As we do this the universal scaling hypothesis breaks at a point predicted by equation (\ref{eqn:p_min}). This is indicated for a single lattice size ($L=11$) as a vertical line.}
\end{figure}

Let us now fix the code distance $L$ and vary the single qubit error rate to see how the full set of data behaves in relation to the universal scaling limit. For each fixed $L$ in Fig. \ref{fig:varyp}, reducing $x$ corresponds to reducing $p$. When $p$ becomes sufficiently small the scaling hypothesis fails and as expected the failure rate deviates below the universal scaling law. 

\begin{figure}[h]
\includegraphics[scale=0.47]{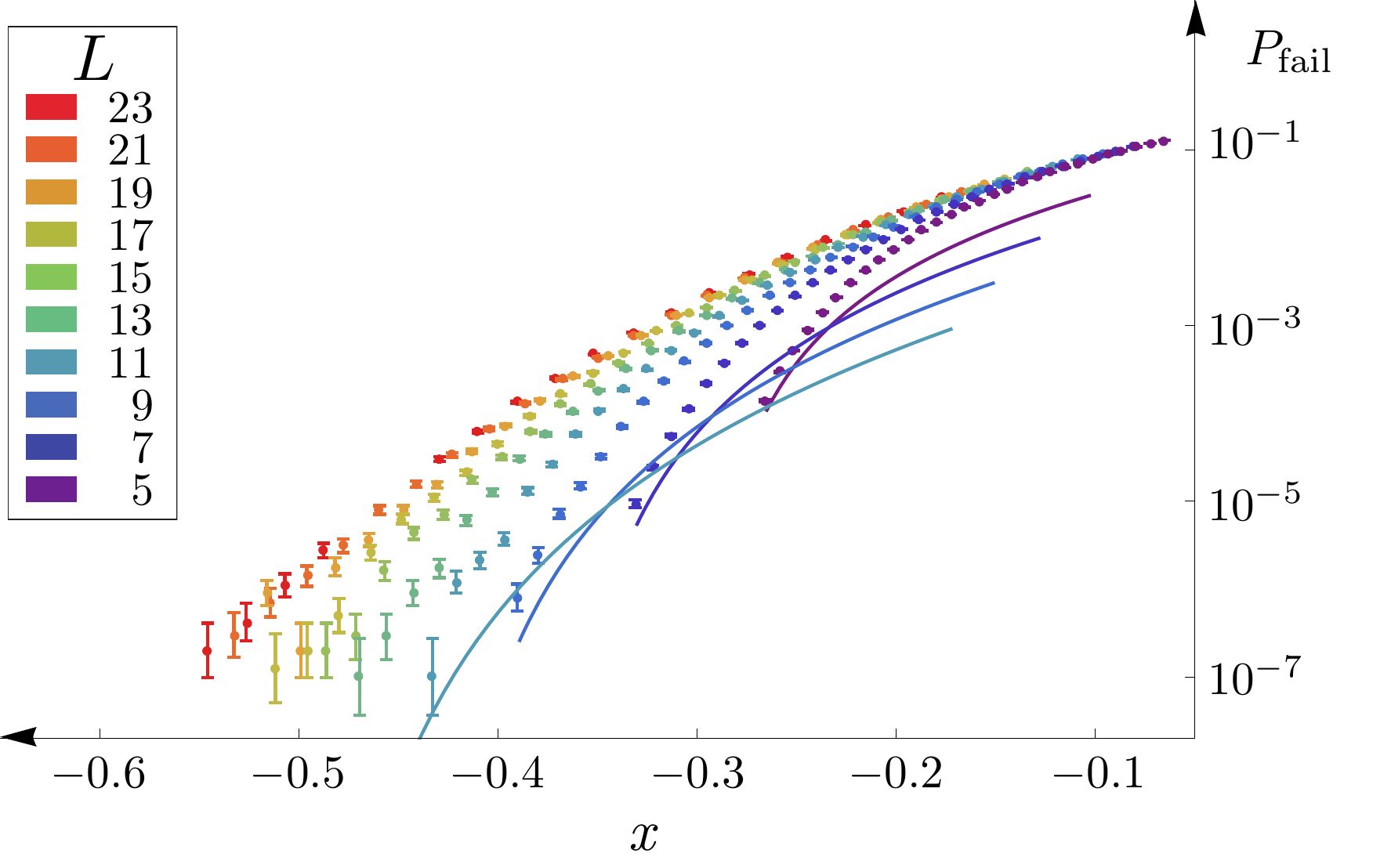} \centering
\caption{\label{fig:low_p}(color online). The full set of renormalized data, colored by lattice size. The low $p$ analytic expression, equation (\ref{eqn:low_p_approx}) is shown for some small lattice sizes. As $x$ decreases the analytic expression tends towards the data. This numerical evidence suggests that the analytic expression is an underestimate of the failure rate for this range of parameters.}
\end{figure}

\subsection{Testing the Range of Validity of the Low Error Rate Regime}

We have proposed that, in the low $p$ limit, spanning errors of the type illustrated in Fig. \ref{fig:logical_fail} dominate when $2L^2 p < \lceil L/2 \rceil$. This is the validity condition we use for the low $p$ regime, see equation (\ref{eqn:p_max}). 

We can rewrite equation (\ref{eqn:low_p_approx}) in terms of $L$ and the rescaling variable, $x$. Fig. \ref{fig:low_p} shows this analytic expression plotted for some small values of $L$ along with the numerical data. As the probability of errors decreases on a fixed lattice the mean number of errors will approach $\lceil L/2\rceil$. As expected, the low $p$ expression gives a good approximation for small lattice sizes and low physical qubit error rates. The data and low $p$ analytic expression converge as $x$ decreases, so for fixed lattice size as the physical error rate decreases the approximation improves. 

\section{Comparison of the overhead in the two regimes}\label{sec:results}

So far we have concentrated on determining the logical error rate as a function of the lattice size and single qubit error rate. Now we wish to demonstrate that it is possible to invert these relationships to find the overhead, $\Omega$. This will be a function of the experimentally determined single qubit error rate, $p$, and maximum tolerable logical failure rate $P_{\mathrm{fail}}$. 

In this work we demonstrate the calculation for the toric code with perfect stabilizer measurements. However the same techniques shown here will also be applicable to more physically realistic settings, for example a planar code with noisy stabilizer measurements. Although the numerics will differ from those presented here, the methods used are expected to be directly analogous.

\begin{figure}
\includegraphics[scale=0.67]{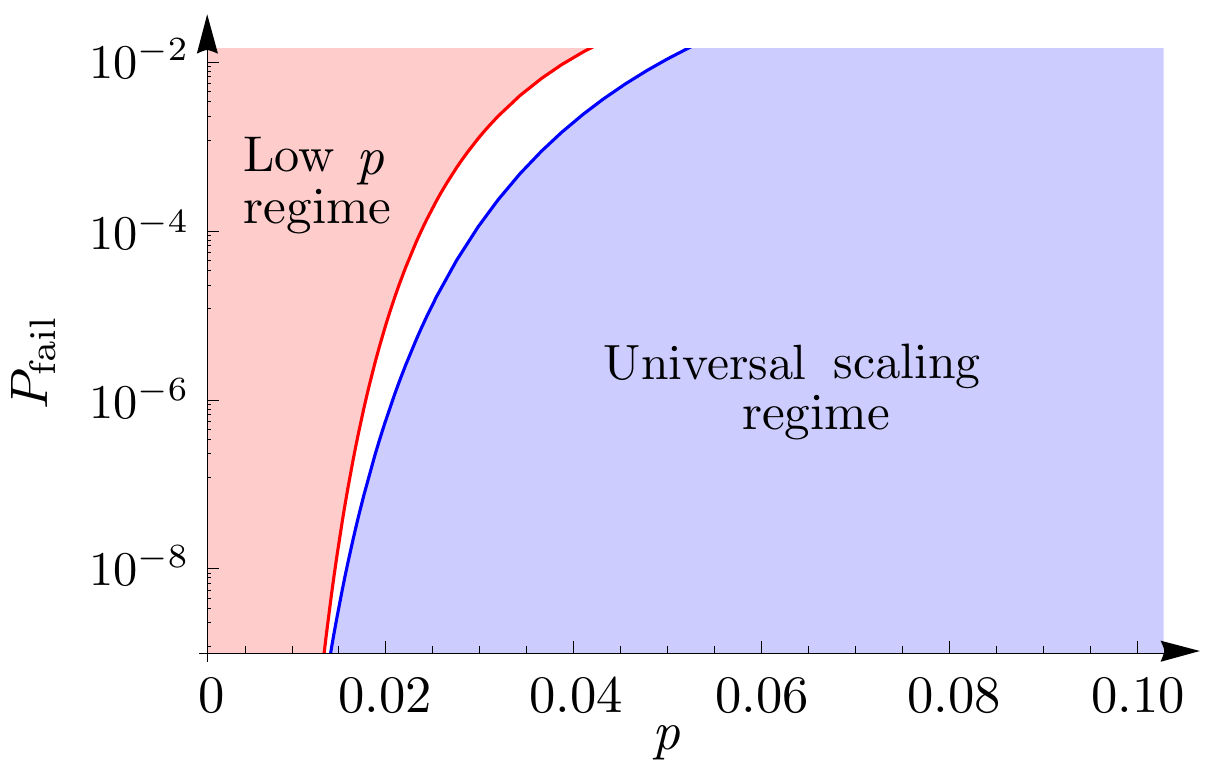} \centering
\caption{\label{fig:validity}(color online). The range of validity of each of the regimes is indicated as a function of the independent variables $p$ and $P_{\mathrm{fail}}$. The uncolored part of the plot is the crossover region between the two regimes.}
\end{figure}

The first step in calculating the overhead is to determine which of the two regimes (universal scaling or low $p$) the code is operating within. To do this we use the expression for $p_{\textsc{ush}}$ in equation (\ref{eqn:p_min}), to find the minimum error rate for which the  universal scaling hypothesis holds. Similarly we find $p_{\textsc{l}p}$, the maximum error rate for which the low $p$ expression holds, using equation (\ref{eqn:p_max}). In Fig. \ref{fig:validity} we plot these two bounds, and the regions of validity that they indicate. Fig. \ref{fig:validity} therefore shows the region of ($P_{\mathrm{fail}}$, $p$) parameter space for which each of the regimes is expected to give a good approximation to the logical error rate. Once the correct regime has been identified, the overhead can be calculated.

In the universal scaling region the logical failure rate is $P_{\mathrm{fail}} = A e^{- a (p-p_{c0})^\nu L}$. By using this to find the lattice size $L$ as a function of $P_{\mathrm{fail}}$ and $p$, and recalling that there are $2L^2$ physical qubits comprising the toric code, we find the overhead in the universal scaling regime is given by:
\begin{eqnarray}
\Omega_{\textsc{ush}}(P_{\mathrm{fail}},p) \:{=}\: \frac{2}{a^2 }\left[\ln \left(-\frac{A}{P_{\mathrm{fail}}}\right) (p-p_{c0})^{-\nu_0 }\right]^2,
\end{eqnarray}
where the constant $a$ has been determined from fits to the data in this work, see Sec. \ref{sec:testing_US_validity}. The remaining parameters, $A$, $p_{c0}$ and $\nu_0$, can be determined from a fit to data generated close to threshold, see Appendix \ref{sec:threshold_calc} for this calculation and for their numerical values.

The analytic expression for the low $p$ regime, equation (\ref{eqn:low_p_approx}), can be simplified by assuming that $\lceil L/2 \rceil = \lfloor L/2 \rfloor = L/2$ and using Stirling's approximation $n! = (n/e)^n \sqrt{2\pi n}$. Inverting this simplified expression we obtain a solution for $L$ that uses the Lambert $W$ function \cite{Corless1996}. We can simplify this using the approximate form for the lower branch of the function \cite{Veberic2010}. It follows that an approximate expression for the overhead in this regime is given by:
\begin{eqnarray}
\Omega_{\textsc{l}p}(P_{\mathrm{fail}},p) \:{=}\: 2 \left[\frac{\ln P_{\mathrm{fail}}^2-\ln\left(-\ln P_{\mathrm{fail}}^2\right)}{\ln 4p }\right]^2.
\label{eqn:low_p_overhead}
\end{eqnarray}

\begin{figure}[h]
\includegraphics[scale=0.48]{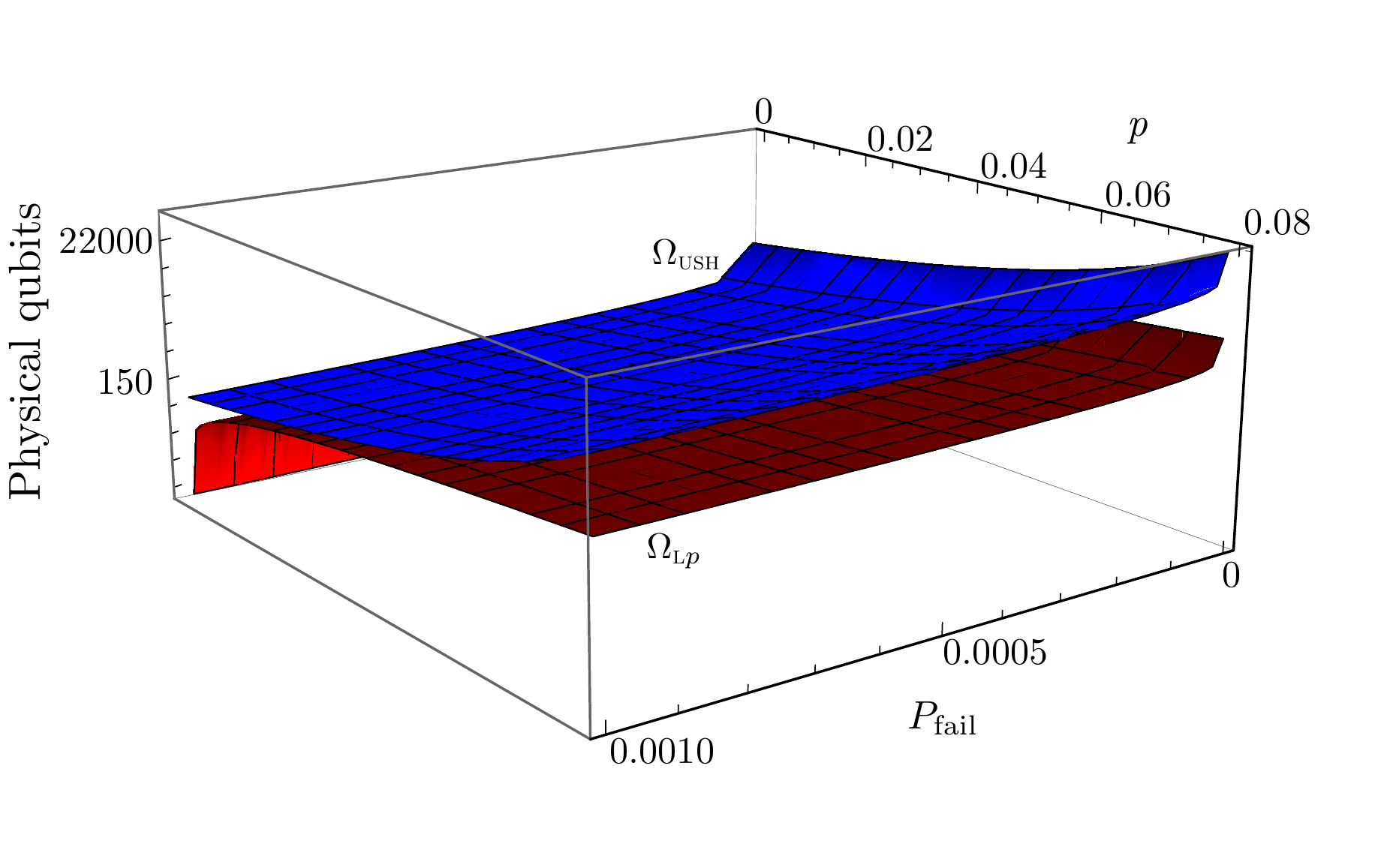} \centering
\caption{\label{fig:oh3d}(color online). A 3-d plot of the overhead, on a logarithmic scale, in each of the two regimes for $0 \le p \le 8\%$ and $10^{-7} \le P_{\mathrm{fail}} \le 10^{-3}$. This plot reveals the gap between the two regimes over the whole region of parameter space considered. It also reveals drop in overhead as the single qubit error rate is reduced, which is particularly striking for the low $p$ regime.}
\end{figure}

\begin{figure}
\begin{flushleft}
(a)
\end{flushleft}
\includegraphics[scale=0.475]{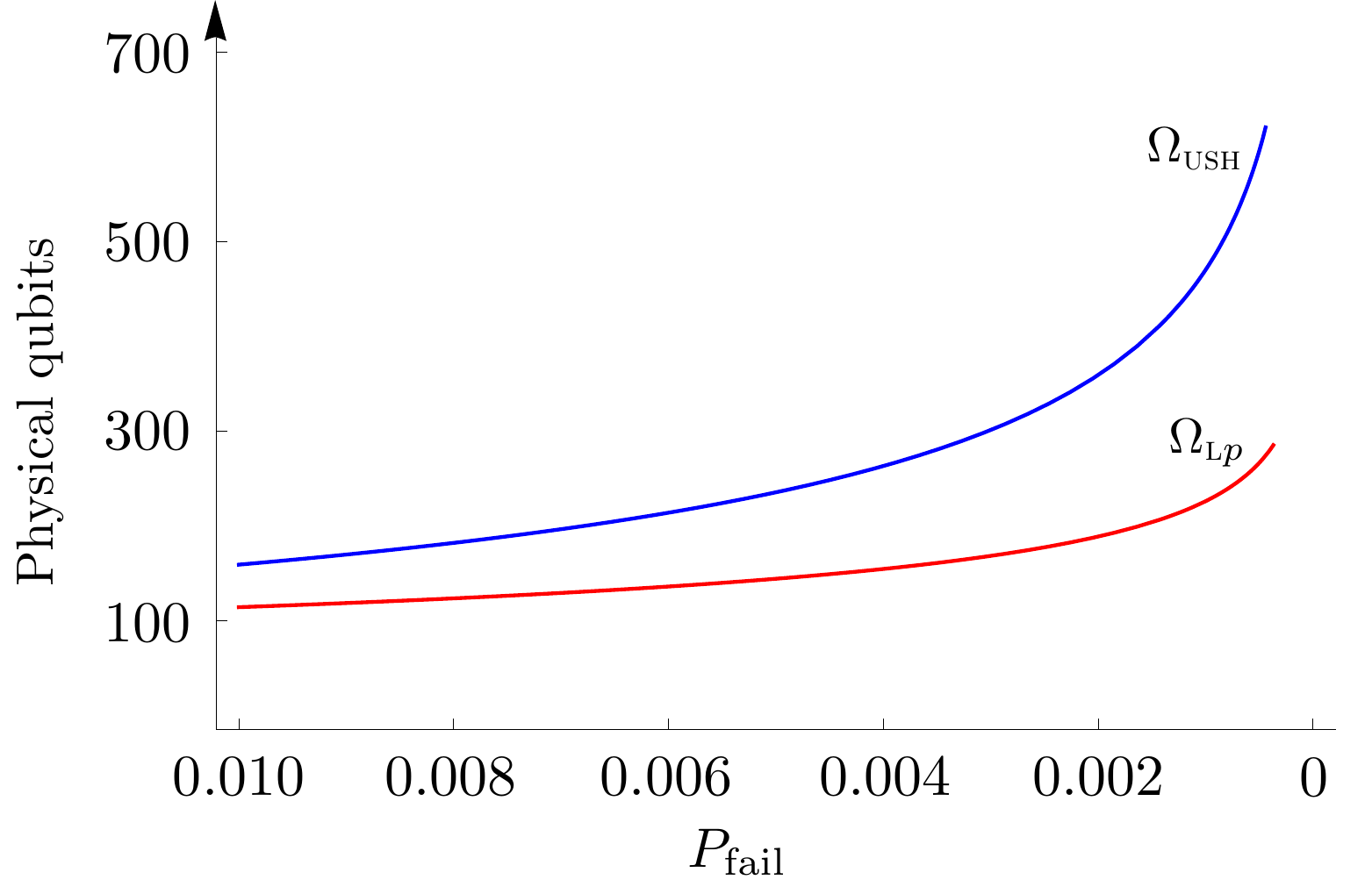} \centering
\begin{flushleft}
\vspace{-2mm}
(b)
\end{flushleft}
\includegraphics[scale=0.48]{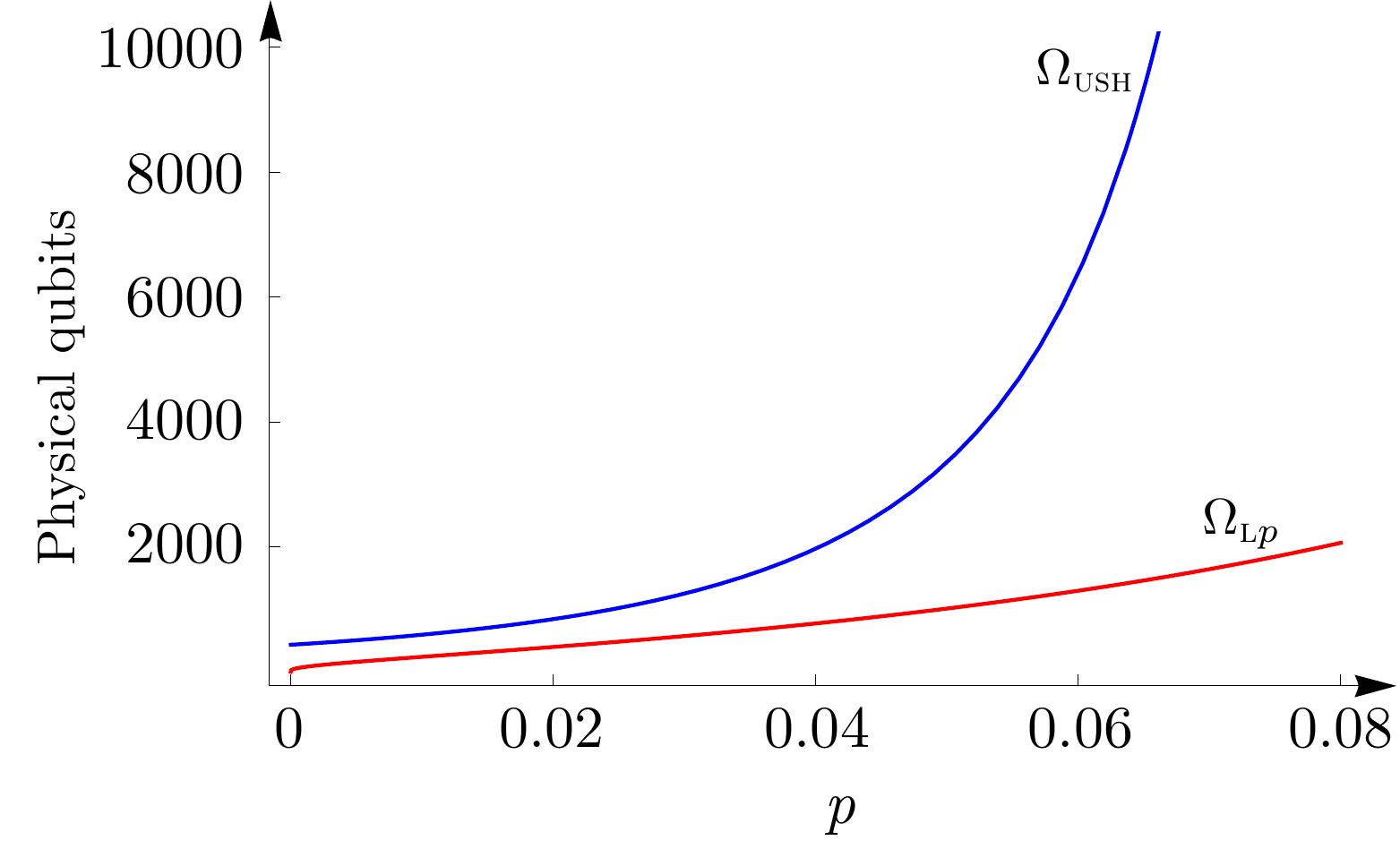} \centering
\vspace{-2mm}
\caption{(color online). (a) The overhead for the toric code calculated for a physical error rate $p=5\%$ for desired fidelities $10^{-7} \le P_{\mathrm{fail}} \le 10^{-2}$. (b) The overhead for logical failure rate $P_{\mathrm{fail}}=10^{-7}$ and $0 \le p \le 8\%$. The plots can be considered to be practical bounds on the overhead for the parameters considered.}
\label{fig:overhead}
\end{figure}

Fig. \ref{fig:oh3d} shows a 3-d plot of the overhead as a function of $P_{\mathrm{fail}}$ and $p$. There is a significant gap between the two plots for most of parameter space (see Fig. \ref{fig:overhead}) and an increase in overhead is seen as both $p$ and $P_{\mathrm{fail}}$ are increased. Allowing a higher logical failure rate will naturally reduce the overhead required, as will reducing the single qubit error probability. 

Fig. \ref{fig:overhead} shows the difference between the required overhead in the two different regimes. For the range of parameters considered the low $p$ expression always gives an estimate of the overhead that lies below the value given by the universal scaling hypothesis.

The low $p$ expression tends to underestimate the logical failure rate for the range of numerical data simulated. Hence this may be considered to be a practical lower bound on the overhead required for those parameters. Conversely, the universal scaling hypothesis is an overestimate of the logical failure rate for most of the numerical data, and hence can be considered to be a practical upper bound to the resources required. 

\section{Conclusions} \label{sec:conclusion}

We have found two distinct operating regimes of the toric code. In one, the data can be rescaled and an ansatz based on this scaling and the exponential dependence of the failure rate on $L$ can be used to find an empirical expression for $P_{\mathrm{fail}}$. In the other, a counting argument gives rise to an analytic expression for the failure rate in the $p \rightarrow 0$ limit. We propose, using the probability distribution of the error weight for fixed $(L,p)$, heuristic conditions for the range of validity of each expression. 

The expressions describing the two regimes have been inverted to calculate the system size required to achieve a desired logical success rate for a given single qubit error rate. We have used the expressions for the logical failure rate to demonstrate techniques to calculate the overhead, $\Omega(P_{\mathrm{fail}}, p)$.  

We expect that the techniques we have demonstrated in this work will be applicable in a wide range of settings. In particular, more physically realistic geometries such as the planar code, whose logical failure rate is expected to higher than that of the toric code \cite{Fowler2013a}. Furthermore, we expect that the methods we have demonstrated can be used to calculate the overhead of a fault-tolerant quantum memory, in which the stabilizer measurements are imperfect. Since all topological codes are based on similar principles the techniques outlined in this work can be expected to be directly applicable despite the fact that the numerics in these cases will differ from those presented here.

Based on the numerical evidence, we claim that for most practical purposes the two regimes bound the required overhead. The numerical results presented in this work are dependent on the choice of the decoder. Similar scaling relationships would be expected for other decoding algorithms, particularly renormalization group-based decoders such as \cite{Duclos-Cianci2010, Bravyi2011}.

This work raises several open questions. It has been shown that the MWPMA decoder has a quadratically lower logical failure rate than the renormalization group algorithm \cite{Fowler2012b}. However, we still believe that a comprehensive comparison of all existing decoders over the whole region of (relevant) parameter space would be interesting and worthwhile. A possible scenario is that the size of the topological code that can be realized will be fixed by technological limitations. In that case, a comparison of the analysis presented in this work for all known decoders below threshold would reveal which should be implemented to minimize the logical failure rate. 

Decoders with high thresholds usually require a longer running time than those with more modest thresholds. We expect a tradeoff between time and space resources, suggesting that those decoders with longer running times may have smaller physical qubit overheads. This is interesting, because although a high threshold is desirable, for practical implementations the running time and physical overhead are also important constraints. Therefore it seems that a balance between these three figures of merit may be of interest for practical quantum computation.

Several of the limitations we faced have been addressed by Bravyi and Vargo in \cite{Bravyi2013} during the preparation of this manuscript. The first of these addresses the crossover region between the two regimes we have identified. Bravyi and Vargo have constructed a heuristic ansatz that interpolates between the dependence on $L$ of the low $p$ regime, $P_{\mathrm{fail}} \propto e^{-\lceil L/2 \rceil}$, and the dependence expected for larger physical error rates, $P_{\mathrm{fail}} \propto e^{-L}$. These functional forms match the two regimes we have identified so the ansatz by Bravyi and Vargo could lead to a method for interpolating between them.

Another benefit of the technique by Bravyi and Vargo is that it provides a fit to the numerical data in the small and moderate $p$ regimes. A significant limitation we faced was the availability of resources to run the Monte Carlo simulations of the error correction procedure. For example, it was impossible to obtain data for $P_{\mathrm{fail}}<10^{-7}$ due to the running time of the decoder. Bravyi and Vargo have discovered a new technique for probing very low error rates on surface codes \cite{Bravyi2013}. Obtaining data for very low logical error rates using this algorithm would help us to verify the conjecture of the range of validity of the low $p$ expression, particularly for larger lattice sizes than we were able to test. 

While heuristic approaches are very flexible, our universal scaling hypothesis has the following advantages. It addresses the large $L$ limit and gives particularly good approximations to the numerical data for moderately large single qubit error rates. The functional form for the universal scaling hypothesis, given in equation (\ref{eqn:US_hypoth}) is derived from the phase transition of the random-bond Ising model, which is a model of statistical physics that the toric code error correction can be mapped to, meaning that it is not a heuristic expression. It is also easily invertible and its pre-factor, $A$, does not depend on the code distance.

Ultimately the implementation of universal quantum computing that is found will set the input parameters that determine which of the regimes it operates within.  

\subsection*{Acknowledgements}

I would like to thank Tom Stace for many valuable discussions in the early stages of this work and his idea of considering universal scaling in such an analysis, as well as for his careful reading of and comments on this manuscript. I would like to thank Dan Browne for his help in preparing this paper, and thank David Jennings and Hussain Anwar for useful discussions and helpful comments on this manuscript. We acknowledge the Imperial College High Performance Computing Service for computational resources. FHEW was supported by EPSRC (grant number: EP/G037043/1).

\appendix

\vspace{5mm}
\section{Determining the threshold}\label{sec:threshold_calc}

In Sec. \ref{sec:US} we rescaled the numerical data using the variable $x = (p-p_{c0})L^{1/\nu_0}$. In order to do this, we must first establish the values of the threshold, $p_{c0}$, and critical exponent, $\nu_0$. The universal scaling hypothesis, equation (\ref{eqn:US_hypoth}), also relies on knowing the failure rate at threshold in the large $L$ limit. In this appendix we show how these quantities are obtained from a fit to data close to the threshold.

The threshold for the stand-alone MWPMA decoding has been calculated previously as $10.306 \pm 0.008 \%$ \cite{Wang2003}. Since we allow the degeneracy of the matching to affect the choice of correction chain, we repeat the calculation in this work to obtain the threshold for our enhanced PMA decoder. 

To find the logical failure rate $P_{\mathrm{fail}}$ we numerically simulate the error correction protocol, enhanced minimum-weight perfect matching (PMA), using the same method described in Sec. \ref{sec:exp_L}.  We performed $N=10^6$ simulations of the error correction procedure for $p$ close to $10.3\%$ and for odd lattice sizes in the range $5 \le L \le 25$. This set of data was only used for the purpose of finding the threshold and critical exponent, and is not the main data set used in this work.

The lattice sizes we use are far from the large $L$ limit, so following the method from Wang {\em et al.} the fitting ansatz was constructed by taking a quadratic expansion in $x$ around the threshold $x=0$ and accounting for finite-size effects by adding a single non-universal term that is dependent on the lattice size \cite{Wang2003}. The ansatz is:
\begin{eqnarray}
P_{\mathrm{fail}} = A + Bx + Cx^2 + DL^{-1/\mu},
\label{eqn:threshfit}
\end{eqnarray}
where $A$, $B$ and $C$ are expansion coefficients, $D$ is the coefficient of the non-universal term, and
\begin{eqnarray}
x = (p - p_{c0})L^{1/\nu_{0}}.
\label{eqn:xdef}
\end{eqnarray}
Here $\nu_0$ is the critical exponent and $p_{c0}$ is the threshold error rate for our PMA decoder. 

Fig. \ref{fig:threshold} shows the rescaled data with finite-size effects subtracted, and the fit to the data. The relevant parameters were found to be:
\begin{eqnarray}
\begin{array}{ccc}
p_{c0} & = & 0.1028 \pm 0.0002, \\
\nu_0 & = & 1.530 \pm 0.006, \\
\mu& = & 1.15 \pm 0.8, \\
A & = & 0.246 \pm 0.006, \\
B & = & 1.87 \pm 0.01,\\
C & = & 2.16 \pm 0.06,\\
D & = & -0.026 \pm 0.008.
\end{array} 
\label{eqn:thresh_values}
\end{eqnarray}

The threshold for our modified decoding algorithms was found to be in agreement with the value found by Wang {\em et al.} for the unmodified MWPMA \cite{Wang2003}. This does not achieve the maximum threshold of $p_{c0} \simeq 10.6\%$ that is possible when the degeneracy of the matching is included \cite{Stace2010}. This is because in the simulations performed for this paper we allow only a weak dependence of the choice of matching on the degeneracy in our modified PMA decoder. This means that the choice of matching is only weakly dependent on the degeneracy of the matching and the effect on the threshold is small. The value of the critical exponent $\nu_0$ found here is in agreement with the value found by Merz and Chalker when calculating the optimal threshold value \cite{Merz2002}, although it does not agree with value found by Wang {\em et al.} for the MWPMA decoder. 

The analysis presented in this appendix establishes the validity of the rescaling approach to the analysis for this choice of decoder by demonstrating that the scaling asatz, equation (\ref{eqn:threshfit}) provides a good fit to the collapsed data close to the threshold.

\vspace{5mm}
\section{Deriving the validity conditions} \label{sec:p_break}

In this appendix we outline the derivation of the validity condition for the universal scaling hypothesis, $p_{\textsc{ush}}$ given in equation (\ref{eqn:p_min}). The validity condition for the low $p$ expression, $p_{\textsc{l}p}$ given in (\ref{eqn:p_max}) is not explicitly shown, but can be reproduced using a similar argument. 

The single qubit errors occur independently and at a rate $p$. The weight of the error that arises, $|E|$, obeys a binomial distribution with a mean that coincides with the typical error weight, 
\begin{eqnarray}
\mu &=& 2L^2 p, \label{eqn:binomial_mean}
\end{eqnarray}
and a variance of: 
\begin{eqnarray}
\sigma^2 &=& 2L^2 p \, (1-p). \label{eqn:binomial_variance}
\end{eqnarray}
According to the central limit theorem the binomial distribution can be approximated by a normal distribution for large enough lattice size. 

For the universal scaling hypothesis, the condition we have proposed is that $\mu$, the mean of the probability distribution, is large with respect to $\lceil L/2 \rceil$. This implies that the weight of the error chain that results is larger than $\lceil L/2 \rceil$ with high probability. We can write this as $\mu \gg \lceil L/2 \rceil$, or 
\begin{eqnarray}
\mu - n\:\frac{\sigma}{2} > \lceil \frac{L}{2} \rceil, 
\label{eqn:USbreak}
\end{eqnarray}
where $n$ is the number of standard deviations above $\lceil L/2 \rceil$ we require the mean to lie. We have chosen $n=2$ for both the universal scaling hypothesis and corresponding condition for the low $p$ expression.

Substituting equations \ref{eqn:binomial_mean} and \ref{eqn:binomial_variance} into equation \ref{eqn:USbreak} we obtain
\begin{eqnarray}
2L^2 p - \sqrt{2L^2 p \, (1-p)} > \lceil \frac{L}{2} \rceil.
\end{eqnarray}
Solving for $p$ and taking only the highest order terms, we arrive at the expression for $p_{\textsc{ush}}$ in equation (\ref{eqn:p_min}).

The expression for $p_{\textsc{l}p}$ in equation (\ref{eqn:p_max}) is obtained similarly, by requiring \begin{eqnarray}
\mu + \sigma < \lceil \frac{L}{2} \rceil.
\end{eqnarray}

\end{document}